\documentclass[twocolumn, aps, prb,showpacs]{revtex4-1}
\usepackage{amsmath}
\usepackage{mathrsfs} 
\usepackage{amssymb} 
\usepackage{graphicx}
\usepackage{subfigure}
\usepackage{dcolumn}
\usepackage{bm}
\usepackage{natbib}

\begin{document}
\title{Probing the uniaxial strains in MoS$_2$ using polarized Raman spectroscopy: A first-principles study}
\date{\today}

\author{Danna Doratotaj}
\affiliation{Department of Physics, Astronomy, and Geosciences, Towson University, 8000 York Road, Towson, MD 21252, USA}
\author{Jeffrey R. Simpson}
\affiliation{Department of Physics, Astronomy, and Geosciences, Towson University, 8000 York Road, Towson, MD 21252, USA}
\author{Jia-An Yan}
\email{jiaanyan@gmail.com}
\affiliation{Department of Physics, Astronomy, and Geosciences, Towson University, 8000 York Road, Towson, MD 21252, USA}

\begin{abstract}
Characterization of strain in two-dimensional (2D) crystals is important for understanding their properties and performance. Using first-principles calculations, we study the effects of uniaxial strain on the Raman-active modes in monolayer MoS$_2$. We show that the in-plane $E'$ mode at 384 cm$^{-1}$ and the out-of-plane $A_1'$ mode at 403 cm$^{-1}$ can serve as fingerprints for the uniaxial strain in this 2D material. Specifically, under a uniaxial strain, the doubly degenerate $E'$ mode splits into two non-degenerate modes: one is $E_{\parallel}'$ mode in which atoms vibrate in parallel to the strain direction, and the other is $E_\perp'$ mode in which atoms vibrate perpendicular to the strain direction. The frequency of the $E_{\parallel}'$ mode blue-shifts for a compressive strain, but red-shifts for a tensile strain. In addition, due to the strain-induced anisotropy in the MoS$_2$ lattice, the polarized Raman spectra of the $E_{\parallel}'$ and $E_{\perp}'$ modes exhibit distinct angular dependence for specific laser polarization setups, allowing for a precise determination of the direction of the uniaxial strain with respect to the crystallographic orientation. Furthermore, we find that the polarized Raman intensity of the $A_1'$ mode also shows evident dependence on the applied strain, providing additional effective clues for determining the direction of the strain even without knowledge of the crystallographic orientation. Thus, polarized Raman spectroscopy offers an efficient non-destructive way to characterize the uniaxial strains in monolayer MoS$_2$.

\end{abstract}

\maketitle

\section{Introduction}

Two-dimensional (2D) materials exhibit a wealth of unusual and fascinating properties, and have become the focus of extensive studies since the discovery of single-layer graphene.\cite{Novoselov2004, Geim2007,Mas2011,Xu2013,Butler2013,Chhowalla2013} Semiconducting transition metal dichalcogenides such as MoS$_2$ have attracted particular attention in recent years.\cite{Chhowalla2013,Wilson1969} Similar to graphene, atomically thin MoS$_2$ layers have been successfully obtained using mechanical exfoliation method.\cite{Mak2010MoS2,Splendiani2010,Radisavljevic2011} Monolayer and few-layer MoS$_2$ exhibit many intriguing physical properties, including a direct optical band gap of about 1.8 eV in the visible range,\cite{Mak2010MoS2,Splendiani2010} strong photoluminescence and electroluminescence,\cite{Mak2010MoS2,Splendiani2010,Cheng2014,Ross2014,Xia2014} and reasonably high mobility.\cite{Baugher2013,Bao2013} Several studies have also shown that monolayer MoS$_2$ exhibits large exciton and trion binding energies,\cite{QiuDY2013,Ye2014,Chernikov2014,He2014,Mak2013} valley selective circular dichroism,\cite{Cao2012,Mak2012,Zeng2012} inversion symmetry breaking together with spin-orbit coupling,\cite{Xiao2012} valley polarization,\cite{Jones2013} and valley Hall effect.\cite{Mak2014} These properties make MoS$_2$ promising for next-generation nano-electronics,\cite{Radisavljevic2011,Wang2012} photonics,\cite{Sanchez2013} photovoltaics,\cite{FengJ2012} and valleytronics.\cite{Cao2012, Mak2012, Zeng2012} In fact, field-effect transistors based on MoS$_2$ have shown a room-temperature electron mobility close to that of graphene nanoribbons, with a current on/off ratio up to 10$^8$. \cite{Radisavljevic2011,FangH2013,KimS2012,FangH2012,Lembke2012,Chuang2014} A CMOS inverter using few-layer phosphorene as the p-channel and MoS$_2$ as the n-channel was recently demonstrated.\cite{LiuH2014} Building blocks of digital circuits containing MoS$_2$, such as logic gates, static random access memory devices, and ring oscillators have been realized.\cite{FangH2013,WangH2012,Radisavljevic5-2011}

Samples of 2D material prepared by exfoliation or epitaxial growth are known to contain inhomogeneous strains that can significantly affect their performance in nanodevices.\cite{FengJ2012, Fischetti2002, Jacobsen2006,Mohiuddin2009,FeiR2014} Determining the magnitude and spatial direction of these strains is important in recognizing the effects of the strain and in implementing strain engineering. Strain engineering has been shown by several studies to be an effective approach to tune properties of nanomaterials for applications.\cite{FengJ2012,Fischetti2002,Jacobsen2006, Mohiuddin2009, FeiR2014,Wang2013, Conley2013,Lu2012,Scalise2011,Johari2012}. It is particularly effective for 2D crystals such as Mo$S_2$ because these materials can sustain much larger strains to even greater than 11\%.\cite{Conley2013,Bertolazzi2011,KimKS2009} It has been demonstrated that the electronic energy band gap of monolayer MoS$_2$ can be further tuned by applying strains.\cite{FengJ2012,Conley2013} Methods of strain mapping are hence important in the study of semiconductors and their applications to nanodevices.

The most common methods used to visualize strain are based on electron microscopy, X-ray scattering or neutron diffraction.\cite{Huber2009} These techniques have disadvantages that they either can alter the sample or have limited spatial resolution. Non-invasive strain mapping is realized by methods such as fluorescence, infrared, or Raman spectroscopy. \cite{Huber2009,Becker2007} In particular, Raman spectroscopy provides an excellent way to study characteristics of materials in a non-destructive way.\cite{Hartschuh2003,Graf2007,Malard2009,HuangM2009,Ferrari2013,LiSL2012,Berkdemir2013,Zhao2013,Luo2013,ZhaoY2013,Tonndorf2013,Ribeiro-Soares2014,Mitioglu2014,YanRusen2014,Terrones2014,Puretzky2015} Monolayer MoS$_2$ is characterized by two Raman-active modes: the in-plane $E'$ mode at 384 cm$^{-1}$ and the out-of-plane $A_1'$ mode at 403 cm$^{-1}$.\cite{LiH2012,LeeC2010,YuY2013,LeeYH2013,Lanzillo2013,Luo2013-2}

Distortions on the lattice change the restoring forces between atoms, and thus affect the vibrational mode frequencies. By probing the changes in the vibrational modes through Raman spectroscopy, it is possible to tell whether a compressive or tensile strain is present. However, more detailed knowledge, such as directions of the strain, is not easy to infer simply from the frequency shift. This information could be useful to understand delicate experimental data such as valley selective circular dichroism.\cite{Cao2012, Mak2012, Zeng2012}

Polarized Raman spectroscopy has recently been applied to investigate 2D materials, including crystalline orientations in ReS$_2$ and strained graphene,\cite{Chenet2015,HuangM2009} anisotropy of black phosphorus,\cite{LingXi2015} and dichroism of helical change in the light in MoS$_2$.\cite{ChenSY2015} In these materials, the Raman intensities of specific modes are sensitive to the crystalline anisotropy, which can be probed through a certain laser polarization setup, i.e., the relationship between polarizations of the incident and scattered laser lights. It is therefore natural to expect that polarized Raman spectroscopy could also be useful to provide information about the uniaxial strain directions in a lattice, since this strain induces a small anisotropy between directions parallel and perpendicular to the strain.

In this work, using first-principles calculations and analytic derivations, we present a detailed study of the relationship between the polarized Raman spectra and the applied uniaxial strain in monolayer MoS$_2$. We find that the uniaxial strains applied in an arbitrary direction induce splitting of the in-plane $E'$ mode into two non-degenerate modes: One is denoted by $E_{\parallel}'$, in which atoms vibrate parallel to the strain direction; the other is denoted by $E_{\perp}'$, where atoms vibrate perpendicular to the strain direction. The $E_{\parallel}'$ mode is very sensitive to the strain, and blue-shifts (red-shifts) for compressive (tensile) strain. Most interestingly, distinct from the doubly degenerate $E'$ mode, the $E_{\parallel}'$ and $E_{\perp}'$ modes now exhibit distinct dependence on the laser polarization angle, thus providing an unambiguous way to characterize the direction of the uniaxial strain. By analyzing the angular dependence of the polarized Raman intensity and the frequency shift, we can uniquely determine both the magnitude and the direction of the uniaxial strain with respect to the crystallographic orientation. Furthermore, we find that the $A_1'$ mode shows very interesting evolution of its intensity for specific polarization setups that offers a unique way to determine the direction of an arbitrary strain even without knowledge of the crystallographic orientation.

The paper is organized as follows. In Section II, we will present the calculation details used in this work. In Section III, results and discussions are presented. We conclude our paper in Section IV.

\section{Calculation Details and Methods}

\begin{figure}[tbp]
  \centering
  \includegraphics[width=0.4\textwidth,clip]{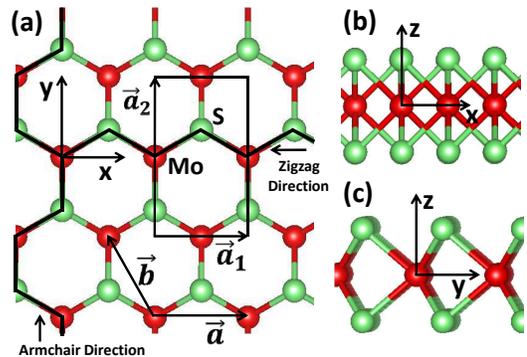}
  \caption{(Color online) (a) Top view and (b) and (c) side views of monolayer MoS$_2$. The $x$, $y$ and $z$ axes have been shown in each case. $\vec{a}$ and $\vec{b}$ are the primitive lattice vectors of MoS$_2$, while $\vec{a}_1$ and $\vec{a}_2$ are the lattice vectors of rectangular supercell adopted to study the uniaxial strain. In this way, after the application of the uniaxial strains $\varepsilon_x$ or $\varepsilon_y$, the supercell can be fully relaxed along another direction. Finally, the crystal structure with the desired strain is obtained. From (a), the atoms form a zigzag atomic chain along the $x$-direction, and an armchair atomic chain along the $y$-direction. }\label{fig1}
\end{figure}

Our calculations were performed using density functional theory (DFT) \cite{Hohenberg64,Kohn65} and density functional perturbation theory as implemented in the Quantum ESPRESSO (QE) code.\cite{pwscf} Local density approximation (LDA) in the Perdew-Zunger scheme was adopted and norm-conserving pseudopotentials were employed to describe the core-valence interactions. The Kohn-Sham equation was solved using a plane-wave cutoff energy of 70 Ry. The $k$-point sampling was set to be a shifted 12$\times$12$\times$1 grid. As shown in Fig.~\ref{fig1}, the atomic plane of Mo atoms is covalently bonded between two planes of S atoms in a trigonal prismatic structure.\cite{Verble1970} From the top view, we can see that the Mo and S atoms are arranged in a hexagonal lattice.\cite{Verble1970} The primitive vectors are $\vec{a}$ = ($a_0$, 0, 0), $\vec{b}$ = (-$a_0$/2, $\sqrt{3}a_0$/2, 0), and $\vec{c}$ =(0, 0, $c$). To simulate the monolayer, we took $c$ = 10$a_0$.

The unit cell of MoS$_2$ was optimized using a 2D variable-cell relaxation method, in which both the cell dimensions and atomic positions were allowed to fully relax. Our optimizations yielded a hexagonal lattice with the lattice constant $a_0$ = 3.134 \AA~ for monolayer MoS$_2$, slightly smaller than the experimental value of 3.15 \AA~ for bulk MoS$_2$.\cite{LeeC2010} This is also in good agreement with previous LDA result of 3.125~\AA.\cite{Sanchez2011} LDA tends to overestimate covalent binding, and therefore it slightly underestimates the lattice constant.

\begin{figure}[tbp]
  \centering
  \includegraphics[width=0.4\textwidth,clip]{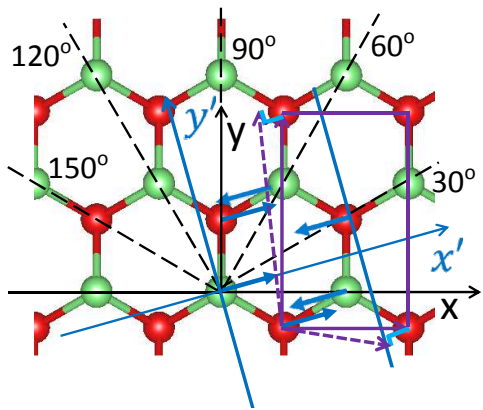}
  \caption{(Color online) The application of strain along an arbitrary direction for monolayer MoS$_2$. The rotated Cartesian frame of reference with the $x'$-axis along the strain direction is shown in blue. The rectangular supercell vectors are shown in purple. After the application of strain in the $x'$-direction, and allowing the relaxation along the $y'$-direction, the final optimized lattice vectors are obtained (purple dashed arrows). The changes in the vector positions are exaggerated, and are much smaller with the 1\% strain that we applied. The symmetry axes along $0^\circ$, $30^\circ$, $60^\circ$, $90^\circ$, $120^\circ$ and $150^\circ$ are shown. The configuration of the atoms is zigzag along the $0^\circ$, $60^\circ$, and $120^\circ$ directions and is armchair along the $30^\circ$, $90^\circ$, and $150^\circ$ directions.}\label{fig2}
\end{figure}

To simulate the MoS$_2$ lattice with a uniaxial in-plane strain, we first chose a rectangular unit cell as shown in Fig.~\ref{fig1}. The uniaxial strain $\varepsilon_x$ or $\varepsilon_y$ was applied along $x$- or $y$-direction, respectively, by changing the corresponding lattice vector to satisfy the following relations: $\varepsilon_x$ = $\frac{a_1'-a_1}{a_1}$, and $\varepsilon_y$ = $\frac{a_2'-a_2}{a_2}$. For strains along arbitrary directions, we changed the reference frame to a rotated Cartesian coordinate system with the $x'$-axis along the strain direction, chose a similar supercell on the crystal lattice, applied the strain along the chosen arbitrary direction, calculated the relaxed structure, and then returned to the reference frame of the original Cartesian coordinates for the subsequent calculations (as shown in Fig.~\ref{fig2}). Following the definition above, a negative $\varepsilon$ denotes a compressive strain, while a positive value implies a tensile strain. When a uniaxial strain was applied along one direction, the unit cell vector along the perpendicular direction was allowed to fully relax until the corresponding stress was below 0.5 kbar and the forces acting on atoms were smaller than 0.02 eV/\AA. For example, with each uniaxial strain applied in the $x$-direction, we applied the resulting $x$ constraint to the structure, and then used a variable-cell relaxation with the degree of freedom only in the $y$-direction. We allowed the structure to completely relax with the necessary change in the $y$-direction cell dimension. Then, based on the obtained cell dimensions and atomic positions, we fixed the cell-dimensions with only the atomic positions being allowed to move to acquire the most energetically favorable positions. The final stress data listed in Table~\ref{tab1} confirmed excellent results, with the final calculated stress only in $x$-direction. For each strain in the $y$-direction, similar calculations were carried out to determine the exact cell dimensions and atomic positions of the final strained structure, again with excellent stress results, as listed in Table~\ref{tab1}. After the structure was fully relaxed, we calculated the vibrational modes at the Brillouin zone (BZ) center ($q= \Gamma$ point) using density-functional perturbation theory as implemented in QE.

\begin{table}
\centering
 \caption{The stresses on MoS$_2$ lattice after relaxation.} \label{tab1}
\begin{ruledtabular}
\begin{tabular}{cccc}
  Direction of  & \begin{large}\textbf{$\varepsilon$}\end{large}  & \multicolumn{2}{c}{Stress after relaxation (kbar)} \\\cline{3-4}
  the Strain &     &  $x$-component  &  $y$-component  \\\hline
  Intrinsic & 0 & -0.48 & -0.48 \\
  \hline
  $x$ & 0.02 & -8.11 & 0.05 \\
    & 0.01 & -4.45 & -0.49 \\
    & -0.01 & 4.16 & 0.39 \\
    & -0.02 & 8.88 & 0.35 \\
\hline
  $y$ & 0.02 & -0.17 & -8.40 \\
   & 0.01 & -0.48 & -4.51 \\
   & -0.01 & 0.35 & 4.04 \\
   & -0.02 & 0.34 & 8.62 \\
\end{tabular}
\end{ruledtabular}
\end{table}

The Raman susceptibility tensor for each mode was obtained by calculating the first derivative of the polarizability tensor and the dynamic matrix. The equation for the Raman intensity of the $j$-th phonon mode is given by: \cite{LiangL2014,Umari2001}

\begin{equation} \label{eq1}
\frac{d\sigma}{d\Omega}=\emph{N}_{prim}\frac{{\omega_{s}}^{4}}{\emph{c}^{4}\emph{V}_{prim}}|\hat{g}_s \cdot \overline{\overline{\alpha}}(\emph{j})\cdot \hat{g}_{i}^{T}|^2\times\frac{\hbar}{2  \omega_{\emph{j}}}(\emph{n}_{\emph{j}}+1),
\end{equation}
where $\omega_{s}$ is the frequency of the scattered photon and $\omega_{j}$ is the frequency of the $j$-th phonon mode of the crystal. Due to energy conservation, $\omega_{s}=\omega_{i}\mp \omega_{j}$, where $\omega_i$ is the incident photon frequency, and the minus (plus) sign applies to the Stokes (anti-Stokes) process. $V_{prim}$ is the volume of the primitive unit cell, $N_{prim}$ is the number of primitive unit cells in the sample, and \emph{c} is the speed of light. $\emph{n}_{\emph{j}}=(\emph{e}^{\hbar\omega_{j}/\emph{k}_{B}\emph{T}}-1)^{-1}$ is the Bose factor of the \emph{j}-th phonon mode. $\hat{g}_{i}$ and $\hat{g}_{s}$ are the polarization unit vectors of the incident and scattered light, respectively. $\overline{\overline{\alpha}}(j)$ is the Raman susceptibility tensor.

Based on Eq.~(\ref{eq1}), we can write the following proportionality relation for the Raman intensity of the \emph{j}-th mode:
\begin{equation}\label{eq2}
I(j) \propto |\hat{g}_s \cdot \overline{\overline{\alpha}}(j)\cdot \hat{g}_i^T|^2,
\end{equation}
where $\hat{g}_i$ and $\hat{g}_s$ are the polarization unit vectors of the incoming and scattered lights, respectively. The components of the $(3\times3)$ symmetric Raman susceptibility tensor $\overline{\overline{\alpha}}(j)$ associated with the $j$-th phonon mode are computed as:\cite{LiangL2014,Umari2001}
\begin{equation}\label{eq3}
\alpha_{\alpha\beta}(j) =V_{prim} \sum_{\mu=1}^N \sum_{l=1}^3 \frac{\partial \chi_{\alpha \beta}}{\partial r_l (\mu)} \frac{\hat{e}_l^j (\mu)}{\sqrt{M_\mu}},
\end{equation}
where $\chi_{\alpha \beta}$ = $(\varepsilon_{\alpha\beta}-\delta_{\alpha\beta})/4\pi$ is the electric polarizability tensor related to the tensor of dielectric constant $\varepsilon_{\alpha\beta}$, $r_l(\mu)$ is the position of the $\mu$-th atom along the direction $l$, $\frac{\partial \chi_{\alpha\beta}}{\partial r_l (\mu)}$ is the first derivative of the polarizability tensor over the atomic displacement, $\hat{e}_l^j (\mu)$ is the displacement of the $\mu$-th atom along the direction $l$ in the $j$-th phonon mode, and $M_\mu$ is the mass of the $\mu$-th atom.\cite{LiangL2014} From the dynamic matrix at the BZ center, $\hat{e}^j$ and $\omega_j^2$ are the calculated eigenvectors and eigenvalues of the matrix, respectively.

The Raman tensor was calculated according to Eq.~(\ref{eq3}) based on the dynamical matrix and the electric polarizability tensor output from QE. Then, for each Raman-active mode, the Raman intensity was analyzed for a given laser polarization setup $(\hat{g}_i$, $\hat{g}_s)$ and laser light wavelength, which finally yielded a Raman spectrum after broadening. Since we are only concerned about the relative intensities, the same laser light wavelength was assumed.

\begin{figure}[tbp]
 \centering
  \includegraphics[angle=-90,width=0.5\textwidth,clip]{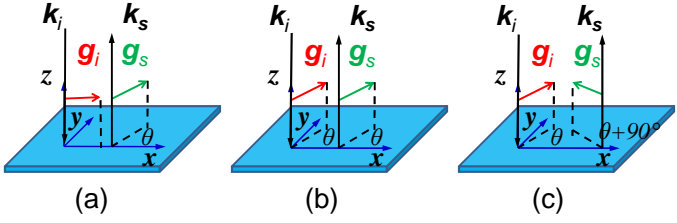}
  \caption{(Color online) Three laser polarization setups considered in this work. Monolayer MoS$_2$ is put on the $xy$-plane as shown in Fig.~\ref{fig3}. The wave vector of the incident laser light is $\vec{k}_i$ =(0, 0, $\bar{1}$), while that of the detected scattered light is $\vec{k}_s$ =(0, 0, 1). (a) $\hat{g}_i$ = (1, 0, 0), $\hat{g}_s$ = ($\cos \theta$, $\sin \theta$, 0). (b) Parallel setup: $\hat{g}_i = \hat{g}_s = (\cos \theta, \sin \theta, 0)$. (c) Cross setup: $\hat{g}_i = (\cos \theta, \sin \theta, 0)$, $\hat{g}_s  = (\cos (\theta+90^\circ), \sin (\theta+90^\circ), 0)$.}\label{fig3}
\end{figure}

Figure~\ref{fig3} shows three typical laser polarization setups. In Fig.~\ref{fig3}(a), the incident laser light is polarized along $x$-direction: $\hat{g}_i$  = (1, 0, 0). The intensity of the scattered laser light is then probed as a function of angle $\theta$ with respect to the fixed $\hat{g}_i$, i.e., $\hat{g}_s = (\cos \theta, \sin \theta, 0)$. In Fig.~\ref{fig3}(b), the incident and scattered laser lights have the same polarization unit vector, and both of them rotate with the same angle of $\theta$ with respect to $x$-axis. This is the so-called parallel polarization setup, which can also be achieved by rotating the samples around the incident laser beams instead. The third laser setup is the cross polarization, as shown in Fig.~\ref{fig3}(c). In this case, the detected scattered light is always polarized perpendicular to the incident light polarization direction. We focus on laser polarization setup in Fig.~\ref{fig3}(a) first. Then we discuss the results for the other two polarization setups.

\section{Results and Discussions}

\subsection{Raman spectra for unstrained MoS$_2$}

Bulk MoS$_2$ is characterized by two Raman-active modes: doubly degenerate $E^{1}_{2g}$ mode at $\sim$382 cm$^{-1}$ and $A_{1g}$ mode at $\sim$407 cm$^{-1}$.\cite{LeeC2010} Note that these modes belong to $E'$ and $A_{1}'$ representations for systems with odd number of layers, respectively. For even number of layers, they change to $E_{g}$ and $A_{1g}$ symmetries.\cite{Terrones2014} For our monolayer system, these modes are denoted by $E'$ and $A_{1}'$, respectively.

\begin{figure}[tbp]
  \includegraphics[width=0.45\textwidth,clip]{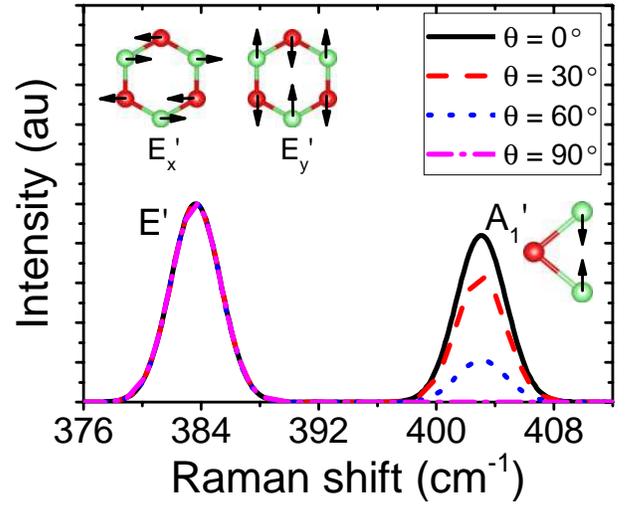}
  \caption{(Color online) Calculated Raman spectra for monolayer MoS$_2$ as a function of $\theta$ for the laser setup shown in Fig.~\ref{fig2}(a). A 2.5 cm$^{-1}$ Gaussian broadening has been applied. The atomic vibration patterns for both $E'$ and $A_1'$ modes are depicted on the insets, respectively.} \label{fig4}
\end{figure}

For the unstrained MoS$_2$ monolayer, there are 6 optical and 3 acoustic phonon modes at the BZ center.  The calculated phonon frequencies for the Raman active $E'$ and $A_1'$ modes are 384 cm$^{-1}$ and 404 cm$^{-1}$, in excellent agreement with the experimental Raman data: $\sim$384 cm$^{-1}$ for the $E'$ mode and $\sim$403 cm$^{-1}$ for the $A_1'$ mode.\cite{LiH2012} The $E'$ mode is doubly degenerate. When the symmetry of the lattice is broken, this mode will split into two non-degenerate modes. For strains applied in $x$- or $y$- directions, these modes can be denoted by $E_x'$ mode (with atoms vibrating along $x$-direction), and $E_y'$ mode (with atoms vibrating along $y$-direction), as shown in the inset of Fig.~\ref{fig4}.

The calculated Raman spectra for monolayer MoS$_2$ are shown in Fig.~\ref{fig4}, using the laser polarization setup as shown in Fig.~\ref{fig3}(a). A small Gaussian broadening has been applied to smear out the Raman peaks. From Fig.~\ref{fig4}, the two peaks corresponding to $E'$ and $A_1'$ modes can be identified clearly. Notably, the intensity for the $E'$ mode is constant and does not depend on the polarization angle $\theta$ between $\hat{g}_s$ and $\hat{g}_i$. In contrast, the intensity of the $A_1'$ mode shows a sensitive dependence on the polarization angle: it starts with the maximum value at $\theta$ = 0$^\circ$, and then decreases with an increasing $\theta$. The intensity of $A_1'$ mode becomes nearly zero for $\theta$ = 90$^\circ$. This result is in good agreement with previous Raman observations by Wang \emph{et al.} \cite{Wang2013}, and with theoretical calculations by Liang \emph{et al.}\cite{LiangL2014}

The dependence of Raman intensity on the polarization angle $\theta$ can be understood based on the Raman tensor for each mode. The calculated Raman tensors for the $E_x'$ and $E_y'$ modes take the following forms, consistent with those reported in the literature:\cite{Loudon1964}
\begin{equation}\label{eq4}
\overline{\overline{\alpha}}({E_x'})=\left(
\begin{array}{ccc}
    0 & a & 0 \\
    a & 0  & 0 \\
    0 & 0 & 0 \\
  \end{array}
\right),
\end{equation}
and
\begin{equation}\label{eq5}
\overline{\overline{\alpha}}({E_y'})=\left(
\begin{array}{ccc}
    a & 0 & 0 \\
    0 & -a  & 0 \\
    0 & 0 & 0 \\
  \end{array}
\right).
\end{equation}
In the unstrained case, the $E'$ mode is doubly degenerate, i.e., $E_x'$ and $E_y'$ modes have the same frequency. Therefore, the calculated intensity will be the summation of those from $E_x'$ and $E_y'$ modes, and the Raman tensor for $E'$ mode reads:
\begin{equation}\label{eq6}
\overline{\overline{\alpha}}({E'})=\left(
\begin{array}{ccc}
    a & a & 0 \\
    a & -a  & 0 \\
    0 & 0 & 0 \\
  \end{array}
\right).
\end{equation}

Based on Eq.~\ref{eq2}, for $\hat{g}_i$ = (1, 0, 0) and $\hat{g}_s$ = (cos$\theta$, sin$\theta$, 0), the intensities of the $E_x'$ and $E_y'$ modes are proportional to $I(E_x')$ $\propto$ $a^{2}\sin^{2} \theta$, and $I(E_y')$ $\propto$ $a^{2}\cos^{2} \theta$, respectively. The total intensity of the $E'$ mode is then $I(E')$ $\propto$ $a^{2}\sin^{2} \theta$+$a^{2}\cos^{2} \theta$ = $a^2$, which is constant and does not depend on $\theta$.

In contrast, $A_1'$ is a non-degenerate mode, and the calculated Raman tensor $\overline{\overline{\alpha}} (A_1')$ in Eq.~(2) takes the well-defined form, again consistent with the literature:\cite{Loudon1964}
\begin{equation} \label{eq7}
\overline{\overline{\alpha}}({A_1'})=\left(
\begin{array}{ccc}
    b & 0 & 0 \\
    0 & b  & 0 \\
    0 & 0 & c \\
  \end{array}
\right).
\end{equation}
Assuming the laser polarization setup with $\hat{g}_i$ = (1, 0, 0) and $\hat{g}_s$ =(cos$\,\theta$, sin$\,\theta$, 0), the Raman intensity of this mode is given by:
\begin{equation} \label{eq8}
I({A_1'}) \propto \bigg | (\cos \, \theta, \sin \, \theta, \, 0) \left(
\begin{array}{ccc}
    b & 0 & 0 \\
    0 & b  & 0 \\
    0 & 0 & c \\
  \end{array}
\right)\left(
\begin{array}{c}
      1 \\
      0 \\
      0\\
      \end{array}\right)\bigg|^2 = b^2 \cos^2\theta,
\end{equation}
\\
which strongly depends on the polarization angle. In particular, the intensity reaches maximum when $\theta$ = 0$^\circ$ and becomes minimum when $\theta$ = 90$^\circ$. This analytical result is in good agreement with the DFT calculations as shown in Fig.~\ref{fig4}(b). Our results are also in good agreement with previous work by Liang \emph{et al.}\cite{LiangL2014}

\subsection{Raman spectra for MoS$_2$ with uniaxial strains along $x$ or $y$ axes}

In this section, we turn to the Raman spectra for monolayer MoS$_2$ under various uniaxial strains. The in-plane uniaxial strain breaks the 3-fold symmetry of the crystal lattice of MoS$_2$. Consequently, the degenerate $E'$ mode splits into two different modes. When the strain is applied along $x$ or $y$ axes, the in-plane atomic vibrations are either along $x$ or $y$. Therefore, we denote the two split modes as $E_x'$ and $E_y'$ modes, respectively (see insets of Fig.~\ref{fig4}).

\begin{figure}[tbp]
  \includegraphics[width=0.5\textwidth,clip]{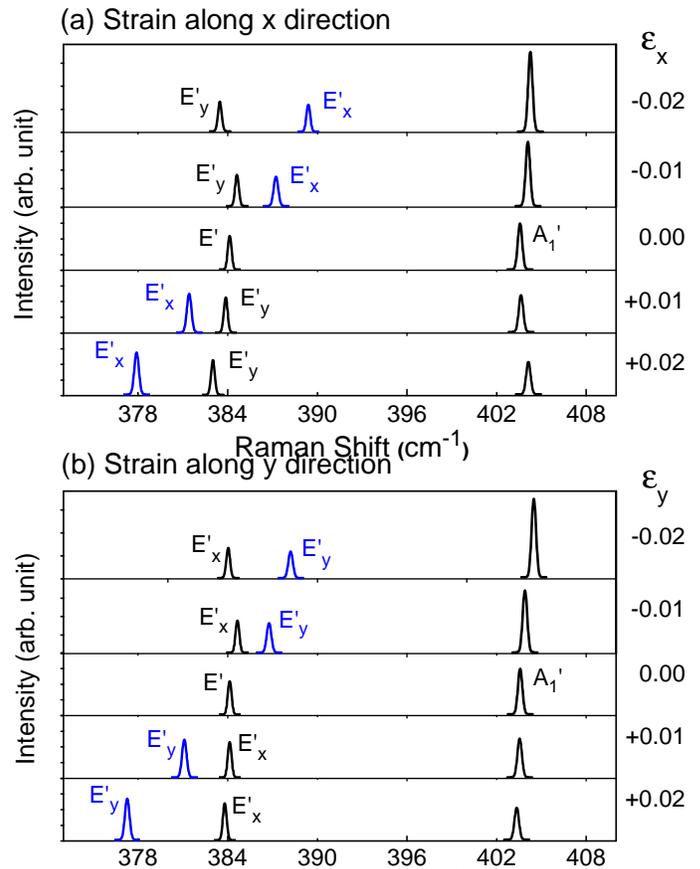}
  \caption{(Color online) Evolution of the mode frequencies and Raman intensities with (a) $\varepsilon_x$ =-2\% to +2\% and (b) $\varepsilon_y$ = -2\% to +2\% in monolayer MoS$_2$. }\label{fig5}
\end{figure}

Figure~\ref{fig5} shows the evolutions of Raman spectra for $E_x'$, $E_y'$, and $A_1'$ modes as a function of the applied strain. Clearly, one in-plane mode blue-shifts for the compressive strain and red-shifts for the tensile strain, while another in-plane mode remains almost unchanged. In contrast, the mode frequency of the $A_1'$ mode is nearly constant for all strains within $\pm$2\%. With the compressive strain applied in the $x$-direction, the $E_x'$ mode frequency increases, because the compression decreases the bond length and therefore increase the restoring forces for the parallel atomic vibrations. In contrast, the tensile strain along $x$ will decrease the restoring forces for the parallel vibrations, leading to a red-shift of the $E_x'$ mode. These results are manifested in Fig.~\ref{fig5}(a). For the strain along $y$-axis, the frequency of the $E_y'$ mode behaves exactly similar to the $E_x'$: It blue-shifts with a compressive strain but red-shifts with a tensile strain, while the $E_x'$ mode frequency remains almost the same, as shown in Fig.~\ref{fig5}(b).

The frequency shift discussed above can be summarized as follows: the mode with atoms vibrating parallel to the strain direction is more sensitive to the strain, and red-shifts (blue-shifts) with a tensile (compressive) strain, while the mode with atoms vibrating perpendicular to the strain direction is almost unchanged. Hereafter, we denote these two modes $E'_\parallel$ and $E'_\perp$, respectively. Previously, these two split modes were denoted by $E'^+$ and $E'^-$ to indicate the upper and lower branches, which essentially correspond to the $E'_\perp$ and $E'_\parallel$ modes, respectively.

From our data, the shift in frequency is 2.8 cm$^{-1}$/\% strain for $E'_\parallel$ mode and is almost zero/\% strain for $E'_\perp$ mode for monolayer MoS$_2$. Experimentally, Conley \emph{et al.}\cite{Conley2013} reported a shift of 4.5 $\pm$ 0.4 cm$^{-1}$/\% strain for the $E'^-$ ($E'_\parallel$) mode, while a shift of 1.0 $\pm$ 0.9 cm$^{-1}$/\% strain for the $E'^+$ ($E'_\perp$) mode for monolayer MoS$_2$. In contrast, Wang \emph{et al.}\cite{Wang2013} showed a shift rate of 2.5 $\pm$ 0.3 cm$^{-1}$/\% strain for the $E'_\parallel$ and 0.8 $\pm$ 0.1 cm$^{-1}$/\% strain for the $E'_\perp$ mode. Our first-principles results are in excellent agreement with their experimental data, as the calculated rate lies within the margins of error in their measurements.

In Table~\ref{tab2}, we present the intensity ratio between the $A_1'$ and the most-shifted $E_x'$ ($E_y'$) modes with the strain applied in $x$ ($y$) direction. Clearly, the ratio shows strong dependence on the magnitude of the strain applied in the lattice: It increases for compressive strains, while decreases for tensile strains. Therefore, in addition to the frequency shift, the intensity ratio between the out-of-plane and the in-plane modes provides an additional clue about the nature of the applied strain.

Comparing Figs.~\ref{fig5}(a) and ~\ref{fig5}(b), the frequency shifts are quite similar to each other, leading to undistinguishable Raman spectra for the uniaxial strain applied in two different directions. Hence, from the Raman frequency shift alone, it is impossible to identify whether the strain is applied along $x$ or $y$ direction. In the following, we explore the method of identifying the strain direction using the polarization dependence of the Raman intensity.

\begin{table}
\centering
\caption{The ratio of the Raman intensities of $A_1'$ to $E_x'$ and $E_y'$ modes versus strain in the $x$ and $y$ directions, respectively.}\label{tab2}
\begin{ruledtabular}
\begin{tabular}{cccc}
  \begin{large}$\varepsilon_{x}$\end{large} & $I(A_1')/I(E_x')$  & \begin{large}$\varepsilon_{y}$\end{large} & $I(A_1')/I(E_y')$ \\
      \hline
  -0.02 & 2.89 & -0.02 & 2.89 \\
  -0.01 & 2.89 & -0.01 & 2.07 \\
  0 & 1.35 & 0 & 1.35 \\
  0.01 & 0.96 & 0.01 & 1.03 \\
  0.02 & 0.78 & 0.02 & 0.78 \\
\end{tabular}
\end{ruledtabular}
\end{table}

\begin{figure}[tbp]
  \centering
  \includegraphics[width=0.5\textwidth,clip]{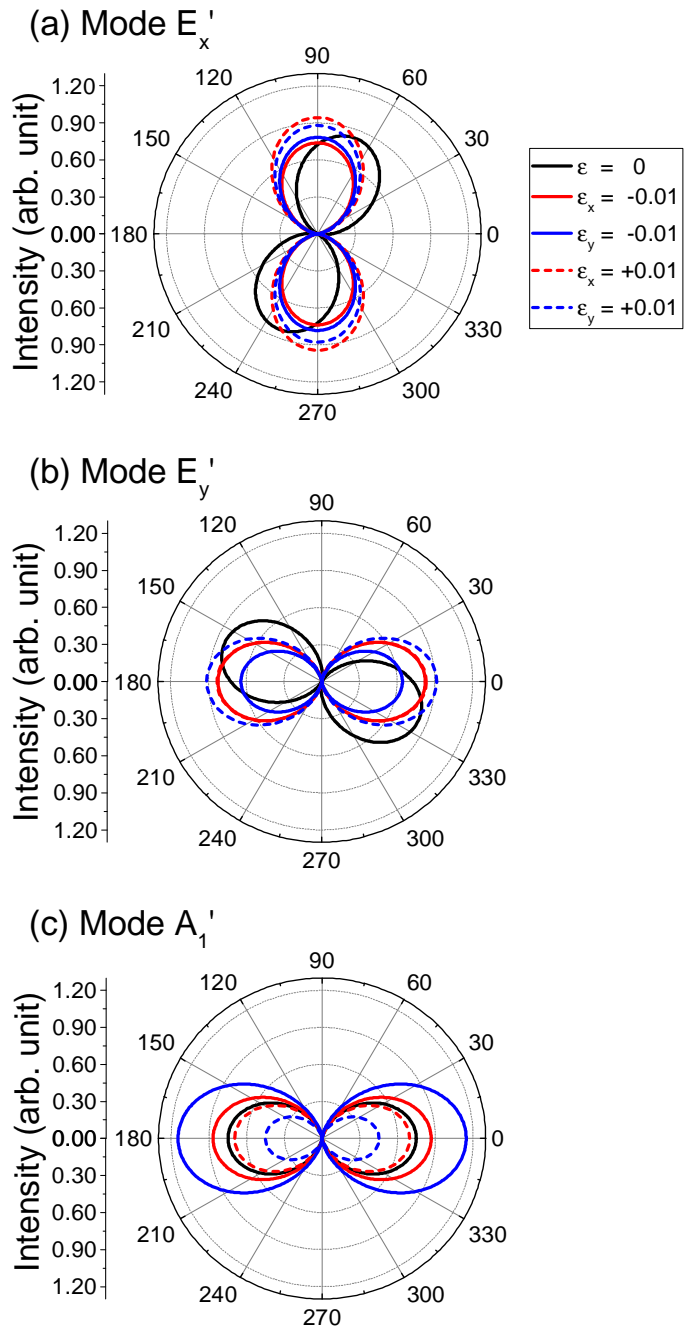}
  \caption{(Color online) The change in the Raman intensity as a function of $\theta$ with respect to the magnitude of strain in the $x$ and $y$ directions. The polarization setup follows Fig.~\ref{fig3}(a). (a) The $E_x'$ mode, (b) the $E_y'$ mode, and (c) the $A_1'$ mode} \label{fig6}
\end{figure}

Figure~\ref{fig6} shows polar plots of the Raman intensity as a function of $\theta$ for $E_x'$, $E_y'$ and $A_1'$ modes, respectively. The polarization setup in Fig.~\ref{fig3}(a) is adopted here. Intensities of both in-plane modes and out-of-plane modes exhibit a dumbbell-shaped dependence on the polarization angle $\theta$. For $\varepsilon$ = 0, the two lobes of $E_x'$ and $E_y'$ modes orient perpendicular to each other, while the Raman scattered light appears to be mainly polarized at angles of about 65$^\circ$ and -25$^\circ$, respectively. In fact, the polarization of scattered light from these two modes could be along arbitrary directions due to the degeneracy of the $E'$ mode. When a strain is applied along $x$ or $y$ directions, the two lobes of the in-plane modes point to $x$- or $y$- directions accordingly. In particular, the two lobes of $E_x'$ mode align along $y$ direction, while the lobes of $E_y'$ mode are along $x$ direction, as shown in Fig.~\ref{fig6}(a) and ~\ref{fig6}(b), respectively. The light scattered by the $E_x'$ mode has the highest intensity in the polarization direction of 90$^\circ$ (along $y$ direction) and the light scattered by the $E_y'$ mode has the highest intensity in the polarization direction of 0$^\circ$ (along $x$ direction). Clearly, the two in-plane modes exhibit distinct angular dependence on the polarization angle. This finding agrees very well with previous Raman work by Wang \emph{et al.}.\cite{Wang2013}

In Fig.~\ref{fig6}(c), the angular dependence of the $A_1'$ mode is somewhat interesting. It shows similar dumbbell structure with the two lobes always pointing along $x$-direction, regardless of the applied strain direction. This is also consistent with the observations for the $A_1'$ mode reported in Ref. [\onlinecite{Wang2013}]. We will discuss further other features of the $A_1'$ mode with different laser polarization setups.

Combining the polarization dependence of the Raman intensity of the two in-plane modes, and the frequency shift as shown in Fig.~\ref{fig5}, it is now possible to distinguish the strain direction. When a compressive strain is applied in the $x$-direction, the $E_x'$ mode will blue-shift from the original frequency, while the frequency of the $E_y'$ mode remains nearly unchanged. This also makes physical sense, as the strain in the $x$-direction should mostly affect the frequency of the phonon mode with vibrations in the $x$-direction. By a similar argument, we reach the conclusion that the mode that is changing frequency with the strain in the $y$-direction is the $E_y'$ mode.

To determine whether the strain is in the $x$- or $y$-direction, we can measure the intensity of the most-shifted Raman mode as a function of the polarization angle, assuming we already know the crystallographic $x$-direction, and the incident laser light is polarized along the $x$-direction. For the laser polarization setup as shown in Fig.~\ref{fig3}(a), the scattered laser light from this mode will polarize mostly perpendicular to the applied strain direction. Namely, when the strain is applied along the $x$ direction, the $E_x'$ mode shifts. Therefore, the polarized scattered laser light will show an angular dependence as shown in Fig.~\ref{fig6}(a). On the other hand, when strain is applied along $y$, the $E_y'$ mode shifts and shows an angular dependence as shown in Fig.~\ref{fig6}(b). This is expected from Eq.~(\ref{eq2}), and can be further explained using the Raman tensors for the $E_x'$ and $E_y'$ modes. Hence, polarized Raman spectroscopy provides a powerful tool to probe the uniaxial strain in MoS$_2$.

Based on Table~\ref{tab2} and Fig.~\ref{fig6}, the magnitude and direction of strain can also be determined by measuring the ratio of the Raman intensity of the $A_1'$ mode to the intensities of $E_x'$ and $E_y'$ modes.  A compression in the $x$-direction causes the intensity of the $A_1'$ mode to increase and the intensity of the $E_x'$ mode to decrease, while a stretch in the $x$-direction results in an opposite change. Similarly, a compression in the $y$-direction causes the intensity of the $A_1'$ mode to increase and the intensity of the $E_y'$ mode to decrease, while a stretch in the $y$-direction has an opposite effect.

\begin{figure*}[tbp]
  \centering
  \includegraphics[width=0.8\textwidth,clip]{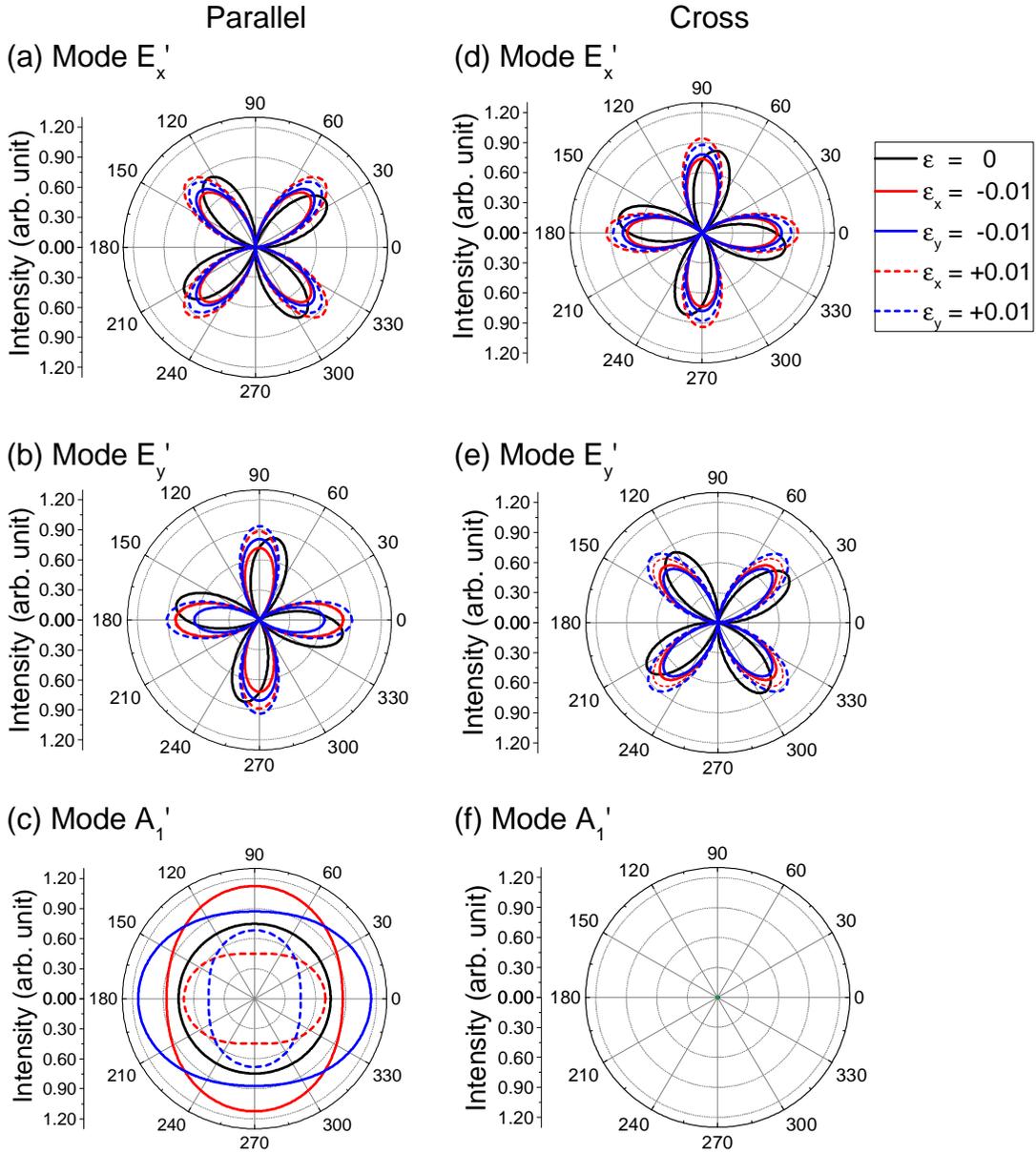}
  \caption{(Color online) The change in the Raman intensity as a function of $\theta$ with respect to the magnitude of strain in the $x$ and $y$ directions for the parallel (left) and cross (right) polarization setups. (a) $E_x'$ mode, (b) $E_y'$ mode, and (c) $A_1'$ mode.}\label{fig7}
\end{figure*}

The above analysis is based on the assumption that we already know the $x$-axis, so that we can fix the polarization unit vector $\hat{g}_i$ along the $x$-direction for monolayer MoS$_2$. In practice, it is difficult to achieve this by using Raman spectroscopy alone. We, therefore, now consider more general laser polarization setups as shown in Figs.~\ref{fig3}(b) and~\ref{fig3}(c). Results are presented in Fig.~\ref{fig7}. Left column is for the parallel polarization setup, while the right column is for the cross polarization setup. The angular dependence of the two in-plane modes shows interesting four-lobe structures with the maximum intensity along directions $n\times$45$^\circ$ ($n$=0-7). In contrast, the out-of-plane $A_1'$ mode exhibits distinct features: the Raman intensity becomes negligible for the cross polarization (see Fig.~\ref{fig7}(f)), while shows an evident dependence on both the nature and direction of the strain applied for the parallel polarization (see Fig.~\ref{fig7}(c)). Such a distinct angular dependence of the out-of-plane $A_1'$ mode provides an even more useful fingerprint for determining the uniaxial strain in MoS$_2$.

The behavior of the angular-dependent Raman intensity for the uniaxial strains along directions of the axes of symmetry ($x$ [zigzag] and $y$ [armchair] directions) can be understood based on their Raman tensors. Consider the $E_y'$ mode first. For the strains applied along $x$ or $y$ or any symmetry axis, the $E'$ mode will split into $E_x'$ and $E_y'$. The Raman tensor of $E_y'$ mode is expressed in Eq.~(\ref{eq5}) above.

Assume the incident laser light is polarized along an arbitrary $\phi$ direction with respect to $x$ axis: $\hat{g}_i$ = (cos$\,\phi$, sin$\,\phi$, 0), the Raman intensity along the polarization direction of $\theta$ will be:
\begin{eqnarray}
I(E_y')\propto\left|\left(
\begin{array}{ccc}
    \cos \,\! \phi & \sin \,\! \phi & 0 \\
  \end{array}
\right)
\left(\begin{array}{ccc}
 a & 0 & 0\\
 0 & -a & 0\\
 0 & 0 & 0\\
\end{array}\right)
\left(
\begin{array}{c}
    \cos \,\! \theta \\
    \sin \,\! \theta \\
    0 \\
  \end{array}
\right)\right|^2\\\nonumber
=(a \, \cos \,\! \phi \, \cos \,\! \theta-a \, \sin \,\! \phi \sin \,\! \theta)^2 = [a \, \cos(\phi+\theta)]^2.
\end{eqnarray}
Clearly, for the incident light polarized in an arbitrary direction $\phi$, the polarization of the scattered light from this mode would be in the $-\phi$ direction. For $\phi \, = \, 0^\circ$, the maximum of intensity will also be along 0$^\circ$, as shown in Fig.~\ref{fig6}(b). Similarly, we can derive the expression for the $E_x'$ mode:
\begin{eqnarray}
I(E_x')\propto\left|\left(
\begin{array}{ccc}
    \cos \,\! \phi & \sin \,\! \phi & 0 \\
  \end{array}
\right)
\left(\begin{array}{ccc}
 0 & a & 0\\
 a & 0 & 0\\
 0 & 0 & 0\\
\end{array}\right)
\left(
\begin{array}{c}
    \cos \,\! \theta \\
    \sin \,\! \theta \\
    0 \\
  \end{array}
\right)\right|^2\\\nonumber
=(a \, \sin \,\! \phi \cos \,\! \theta - a \cos \,\! \phi \sin \,\! \theta)^2 = [a \sin(\phi+\theta)]^2.
\end{eqnarray}
Therefore, for an incident light polarized in an arbitrary direction $\phi$, the maximum intensity of the scattered light for the $E_x'$ mode will be polarized along $\theta \, = \, 90^\circ-\phi$. When $\phi \, = \, 0^\circ$, $\theta \, = \, 90^\circ$, as shown in Fig.~\ref{fig6}(a).

For the parallel setup shown in Fig.~\ref{fig3}(b), $\hat{g_i}$ = $\hat{g_s}$ = (cos$ \, \theta$, sin$ \, \theta$, 0), i.e., $\phi \, = \, \theta$. The Raman intensity of the scattered light by the $E_x'$ and $E_y'$ modes are
$I(E_x')\propto [a \, \cos(2\theta)]^2$, and $I(E_y')\propto [a \, \sin(2\theta)]^2$, respectively. The intensity for the $E_x'$ mode reaches maximum when $\theta \, = \, n\pi/2$, while it reaches maximum for the $E_y'$ modes at $\theta \, = \, n\pi/2 \pm \pi/4$, with $n$ an integer. This explains the four-lobe structures in Fig.~\ref{fig7} (a) and (b). Similarly, we can derive the expressions that explain the behavior of intensity with $\theta$ in Fig.~\ref{fig7} (d) and~\ref{fig7}(e).

For the mode $A_1'$, using Eq.~(\ref{eq2}) and the $A_1'$ Raman tensor in Eq.~(\ref{eq7}), we will have the following expression for the Raman intensity:

\begin{eqnarray}
I(A_1')\propto\left|\left(
\begin{array}{ccc}
    \cos \,\! \phi & \sin \,\! \phi & 0 \\
  \end{array}
\right)
\left(\begin{array}{ccc}
 b & 0 & 0\\
 0 & b & 0\\
 0 & 0 & c\\
\end{array}\right)
\left(
\begin{array}{c}
    \cos \,\! \theta \\
    \sin \,\! \theta \\
    0 \\
  \end{array}
\right)\right|^2\\\nonumber
=(b \, \cos \,\! \phi \, \cos \,\! \theta+b \, \sin \,\! \phi \sin \,\! \theta)^2 = [b \, \cos(\phi-\theta)]^2.
\end{eqnarray}

Obviously, the maximum intensity occurs when $\theta$ = $\phi$. Thus, for an incident light polarized in an arbitrary direction $\phi$, the polarization of the scattered light for mode $A_1'$ will be in the same direction $\phi$.  For $\phi$ = 0$^\circ$, the maximum intensity will be along 0$^\circ$. This explains the result in Fig.~\ref{fig6} (c).

To explain the results in Fig.~\ref{fig7} (c) and \ref{fig7}(f), we note that with the application of the strain in the $x-$ or $y-$ direction, the Raman tensor of the $A_1'$ mode changed to the following form:

\begin{equation} \label{eq12}
\overline{\overline{\alpha}}({A_1'})=\left(
\begin{array}{ccc}
    a & 0 & 0 \\
    0 & b  & 0 \\
    0 & 0 & c \\
  \end{array}
\right),
\end{equation}
with $a\neq b$. For a compressive (tensile) strain in the $x$-direction ($y$-direction), $a$ is larger than $b$, while for a tensile (compressive) strain in the $x$-direction ($y$-direction), $a$ is smaller than $b$. The expression for the intensity of the $A_1'$ mode follows:

\begin{eqnarray}
I(A_1')\propto(a \, \cos \,\! \phi \, \cos \,\! \theta+b \, \sin \,\! \phi \sin \,\! \theta)^2.
\end{eqnarray}

With the polarization setup in Fig.~\ref{fig3}(b), $\theta$ = $\phi$. We thus have:
\begin{equation}\label{eq14}
 I(A_1') \propto (a \,\!  \cos ^{2} \,\! \theta + b \,\! \sin ^{2} \,\! \theta)^{2} \,\! = \,\! [a+(b-a)\sin ^{2} \,\! \theta]^{2}.
\end{equation}
\\
Based on this expression, when $b$ is more than $a$, the maximum intensity occurs at $\theta$ = 0$^\circ$ and the minimum intensity occurs at $\theta$ = 90$^\circ$. When $a$ is more than $b$, we have the opposite. This clearly explains the results in Fig.~\ref{fig7}(c) when the obtained Raman tensors in our data are taken into account.

For the cross polarization setup in Fig.~\ref{fig3}(c), $\theta$ = $\phi$ + 90$^\circ$; therefore, we will have $I(A_1')\propto[a\cos(\theta-90^\circ) \cos\theta  + b\sin(\theta-90^\circ)\sin\theta]^{2} = (a-b)^{2}\sin^{2}(2\theta)/2$. From our data $|a-b|$ was about 10 times smaller that $a$ and $b$. Thus, the intensity with this setup would be smaller than the intensity in Fig.~\ref{fig7}(c) by a factor of about 100. That is why we have the intensities of nearly zero in Fig.~\ref{fig7}(f); however, when we zoomed in on the point at the center, we saw the minuscule four-lobe structures that incorporated the factor $\sin^2(2\theta)$ as expected from this derivation. For all practical purposes, these intensities will not be experimentally detectable and therefore can be considered zero.

Finally, we would like to mention that the two elements of the Raman tensor for both $E_x'$ and $E_y'$ are slightly changed by the strain, showing variations along $x$ and $y$. This introduces a small difference in the maximum intensity along different directions, as shown in Fig.~\ref{fig7}.

\subsection{Raman spectra for MoS$_2$ with uniaxial strains along arbitrary directions}

We now discuss the Raman spectra of MoS$_2$ under a uniaxial strain applied along an arbitrary direction. As shown in Fig.~\ref{fig1}, the monolayer MoS$_2$ crystal lattice possesses point group symmetry of D$_{3h}$.\cite{ZhaoY2013, Ribeiro-Soares2014,LiangL2014,Sanchez2011} The three in-plane axes in the directions of $30^\circ$,  $90^\circ$, and $150^\circ$ with respect to the $x$-axis are equivalent, and can be regarded as along the armchair atomic chain direction. In contrast, the three axes along $0^\circ$, $60^\circ$, and $120^\circ$ are along the zigzag atomic chain direction (see Fig.~\ref{fig2}). The only physically non-equivalent directions are limited to 0$^\circ$ to 30$^\circ$. Below, we focus on the uniaxial strain between $0^\circ$ to $30^\circ$, using parallel laser polarization setup.

\begin{figure*}[tbp]
  \centering
  \includegraphics[angle=-90,width=\textwidth,clip]{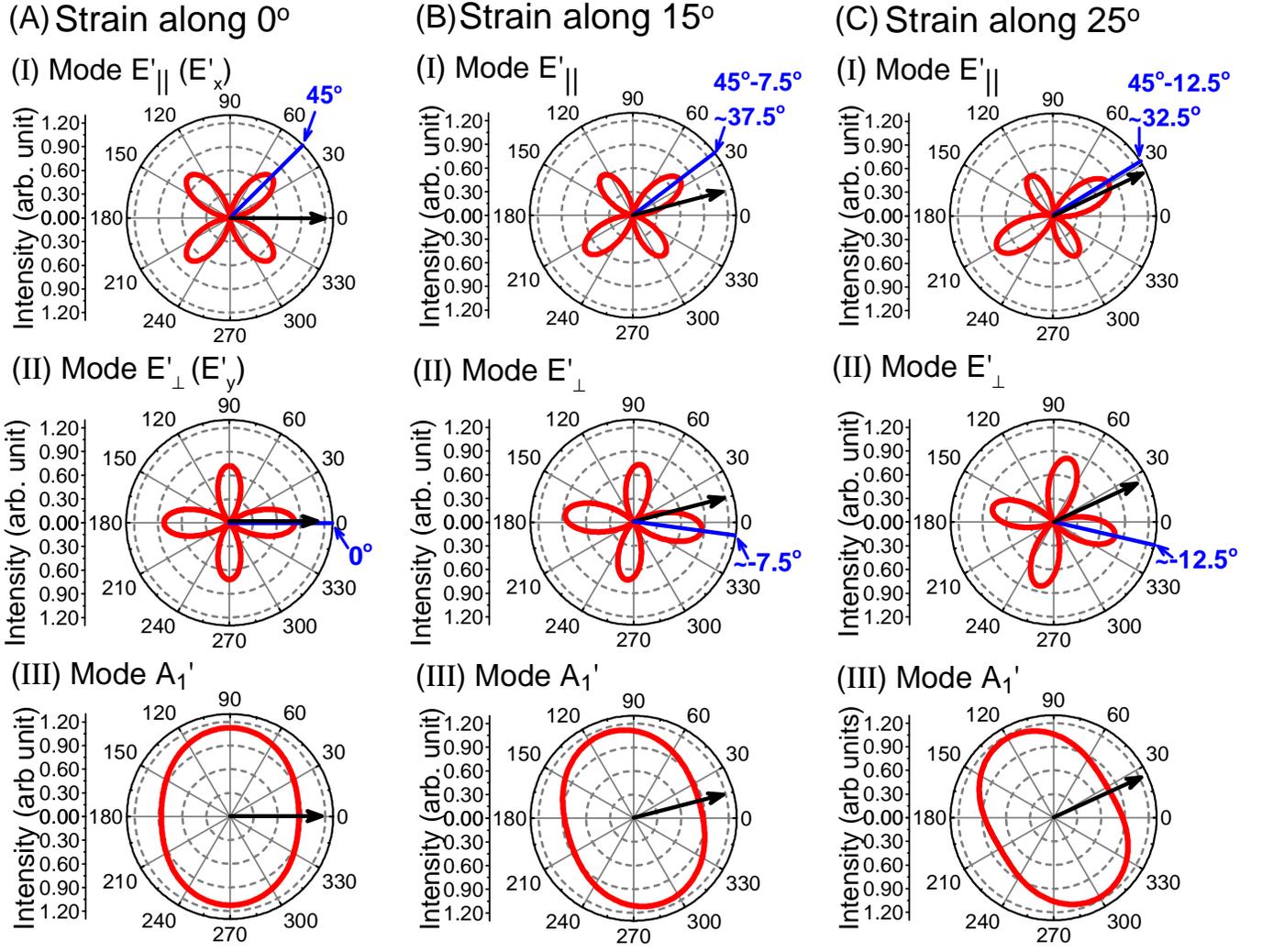}
  \caption{(Color online) The Raman intensity versus angle $\theta$ with the polarization setup $\hat{g_i}$ = $\hat{g_s}$ = (cos $\theta$, sin $\theta$, 0) for 0.01 compression strain in three directions: 0$^\circ$, 15$^\circ$ and 25$^\circ$.}\label{fig8}
\end{figure*}

When strain is applied along any direction between $0^\circ$ and $30^\circ$, the $E'$ in-plane phonon mode will split into two branches, one of which vibrates in the direction parallel to the applied strain direction, and the other vibrates in the direction perpendicular to the strain. This has been shown for graphene in previous work.\cite{Mohiuddin2009} Our calculations for several strain directions, including $0^\circ$, $15^\circ$, $25^\circ$, and $60^\circ$, confirm that this is also true for MoS$_2$. Thus, for a strain in an arbitrary direction, the vibrational modes can no longer be considered pure $E_x'$ and $E_y'$ modes, instead, we will denote them as $E_\parallel'$ and $E_\perp'$ modes.

We apply a 1\% compressive strain in two arbitrary directions: $15^\circ$ and $25^\circ$ with respect to the $x$-axis in Fig.~\ref{fig1}. We consider the parallel polarization setup only (Fig.~\ref{fig3}(c)). The frequency of the $E_\parallel'$ mode blue-shifts for a compressive strain, while red-shifts for a tensile strain, similar to the results discussed in Section III (B).

The Raman intensities of $E_\parallel'$, $E_\perp'$ and $A_1'$ modes versus the angle $\theta$ are depicted in Fig.~\ref{fig8}. The results for the 0.01 compressive strain along $\phi=0^\circ$ are also shown for comparison. Clearly, intensities of the in-plane modes all exhibit similar four-lobe structures as a function of the angle $\theta$, except that there is a small rotation with respect to the $x$-axis for $\phi$ = 15 and 25$^\circ$. These directions have been indicated in Fig.~\ref{fig8}(B) and ~\ref{fig8}(C). For example, the intensity of $E_\parallel'$ mode reaches maximum at angle of 37.5$^\circ$, while that of $E_\perp'$ mode is maximized along -7.5$^\circ$.

To understand the calculation results, we first investigate the Raman susceptibility tensor. The Raman susceptibility tensors for $E_\parallel'$ and $E_\perp'$ modes are found to have the following forms, where $\phi$ is the direction of the strain:
\begin{equation}
\overline{\overline{\alpha}}({E_\perp'})=\left(
\begin{array}{ccc}
    \cos \,\! \phi & \sin \,\! \phi & 0 \\
    -\sin \,\! \phi & \cos \,\! \phi & 0 \\
    0 & 0 & 1 \\
  \end{array}
\right)
\left(\begin{array}{ccc}
 a & 0 & 0\\
 0 & -a & 0\\
 0 & 0 & 0\\
\end{array}\right),
\end{equation}
and
\begin{equation}
\overline{\overline{\alpha}}({E'_\parallel})=\left(
\begin{array}{ccc}
    \cos \,\! \phi & \sin \,\! \phi & 0 \\
    -\sin \,\! \phi & \cos \,\! \phi & 0 \\
    0 & 0 & 1 \\
  \end{array}
\right)
\left(\begin{array}{ccc}
 0 & a & 0\\
 a & 0 & 0\\
 0 & 0 & 0\\
\end{array}\right).
\end{equation}

For the parallel polarization, $\hat{g_i}$ = $\hat{g_s}$ = (cos $\theta$, sin $\theta$, 0). The strain is applied in the arbitrary direction $\phi$ in the $xy$ plane. For $E_\perp'$ mode:
 \begin{eqnarray}\label{eq17}
I(E_\perp') \! \propto \! \nonumber \left| \left(\!
\begin{array}{cc}
    \cos \,\! \theta \!&\! \sin \,\! \theta \\
  \end{array}
\!\right)\!\!
\left(\!\!\!
\begin{array}{cc}
    \cos \,\! \phi & \sin \,\! \phi \\
    -\sin \,\! \phi & \cos \,\! \phi \\
  \end{array}
\!\right)\!\!
\left(\!\begin{array}{cc}
 a & 0 \\
 0 & -a \\
\end{array}\!\right)\!\!
\left(\!
\begin{array}{c}
    \cos \,\! \theta \\
    \sin \,\! \theta \\
  \end{array}
\!\right)\!  \right|^2\!\!\!\!=\\\nonumber
[a \,\! \cos(\theta+\phi) \cos \,\! \theta-a \sin(\theta+\phi) \sin \,\! \theta]^2 \! = [a \,\! \cos(2\theta+\phi)]^2\!.\\
\end{eqnarray}
Clearly, the maximum intensity occurs at $\theta$ = \begin{large}$\frac{-\phi}{2}$\end{large} and 90$^\circ$-\begin{large}$\frac{\phi}{2}$\end{large}.

For $E_\parallel'$ mode:
 \begin{eqnarray}\label{eq18}
I(E_\parallel') \! \propto \! \nonumber\left|\left(\!
\begin{array}{cc}
    \cos \,\! \theta \!&\! \sin \,\! \theta \\
  \end{array}
\!\right)\!\!
\left(\!\!\!
\begin{array}{cc}
    \!\cos \,\! \phi & \sin \,\! \phi \\
    \!-\sin \,\! \phi & \cos \,\! \phi \\
  \end{array}
\!\right)\!\!
\left(\!\begin{array}{cc}
 \!0 & a \\
 \!a & 0 \\
\end{array}\!\right)\!\!
\left(\!
\begin{array}{c}
    \cos \,\! \theta \\
    \sin \,\! \theta \\
  \end{array}
\!\right)\! \right|^2\!\!\!\!=\\\nonumber
[a \,\! \sin(\theta+\phi) \cos \,\! \theta\, + \, a \,\! \cos(\theta+\phi) \sin \,\! \theta]^2 \!= [a \,\! \sin(2\,\!\theta+\phi)]^2\!.\\
\end{eqnarray}
\\[0in]
The maximum intensity occurs at $\theta$ = 45$^\circ$-\begin{large}$\frac{\phi}{2}$\end{large} and 135$^\circ$-\begin{large}$\frac{\phi}{2}$\end{large}, in good agreement with numerical data shown in Fig.~\ref{fig8}.

Going from $0^\circ$ to $30^\circ$, when the strain direction angle $\phi$ reaches $30^\circ$ (which is one of the axes of symmetry of the crystal), the Raman intensity of the $E_\parallel'$ mode will be maximum at $45^\circ-15^\circ=30^\circ$, which is again in the direction of the strain. In this $E_\parallel'$ mode, atoms will vibrate along the zigzag direction. At the start at $0^\circ$, the $E_\parallel'$ mode vibrates along the armchair chain direction.

The above results and analysis for the $E_\perp'$ and $E_\parallel'$ modes are in agreement with the previous work done in graphene,\cite{Mohiuddin2009} and in MoS$_2$\cite{Wang2013}. In Ref.~\onlinecite{Mohiuddin2009}, an effective photon-phonon interaction Hamiltonian has been introduced for the $E_{2g}$ mode:

\begin{equation}\label{eq19}
H_{int} \propto [(\large{\varepsilon}^{i}_x\large{\varepsilon}^{s}_x-\large{\varepsilon}^{i}_y\large{\varepsilon}^{s}_y)u_y-
(\large{\varepsilon}^{i}_x\large{\varepsilon}^{s}_y+\large{\varepsilon}^{i}_y\large{\varepsilon}^{s}_x)u_x],\normalsize
\end{equation}
where $\Large{\varepsilon}^{i(s)}_x,\Large{\varepsilon}^{i(s)}_y$ are the Cartesian components of the electric field of the incident (scattered) light and $\Large{u_x,u_y}$ are the phonon displacements in the $(x,y)$ basis (where $x$ and $y$ are the zigzag and armchair crystallographic directions, respectively). Based on this, the intensities for the two split in-plane phonon modes can also be derived. The expressions are similar for graphene and MoS$_2$ and have the form:\cite{Mohiuddin2009}

\begin{equation}\label{eq20}
I(E'^-) \propto \sin^2(\theta_i+\theta_s+3 \, \phi),
\end{equation}
\begin{equation}\label{eq21}
I(E'^+) \propto \cos^2(\theta_i+\theta_s+3 \, \phi),
\end{equation}
where $\phi$ is the angle of the strain axis with respect to the crystallographic $x$-direction, and $\theta_i$ and $\theta_s$ are the angles of the polarization directions of the incident and scattered light with respect to the strain axis. In our case, $\theta_i$ and $\theta_s$ are measured with respect to the crystallographic $x$-axis, then $3\,\phi$ needs to be replaced with $\phi$ in Eqs.~(\ref{eq20}) and (\ref{eq21}). Replacing $\theta_i$ = $\theta_s$ = $\theta$ in thus modified Eqs.~(\ref{eq20}) and (\ref{eq21}) yields Eqs.~(\ref{eq17}) and (\ref{eq18}), which confirms our data and analysis.

Interestingly, for the $A'_1$ mode, we see that the minimum Raman intensity occurs at the direction of the strain and the maximum intensity occurs at the direction perpendicular to it in all cases, as shown in the last row of Fig.~\ref{fig8}. To further understand this result, based on Eq.~(\ref{eq14}), we expect that the expression for $I(A_1')$ is of the form:
\begin{equation}\label{eq22}
[a+(b-a)\sin ^{2}(\theta-\phi)]^{2}
\end{equation}
for the above result to occur; where $\phi$ is the angle of the direction of the strain with the $x$-axis. We also note in our data that with the strain in the arbitrary direction, the $A_1'$ mode Raman tensor has the following form:
\begin{equation} \label{eq23}
\overline{\overline{\alpha}}({A_1'})_{\phi}=\left(
\begin{array}{ccc}
    a & d & 0 \\
    d & b  & 0 \\
    0 & 0 & c \\
  \end{array}
\right).
\end{equation}
with $d$ nonzero. Then, we ask by what matrix should we multiply the $A_1'$ Raman tensor given in Eq.~(\ref{eq12}) (the case when the strain is along one of the axes of symmetry), such that the resulting matrix would be the Raman tensor in Eq.~(\ref{eq23}).  We did the multiplication of the Eq.~(\ref{eq12}) tensor by an unknown $(3\times3)$ matrix denoted $\mathbb{T}_X$, and put the resulting tensor in Eq.~(\ref{eq2}) (with $\hat{g_i}$ = $\hat{g_s}$ = (cos $\theta$, sin $\theta$, 0) for the setup in Fig.~\ref{fig8}) to obtain the expression for the intensity $I(A_1')$. Then, we expanded Eq.~(\ref{eq22}), and set the like terms equal. Taking into account the observed form of the tensor in Eq.~(\ref{eq23}), this resulted in three equations, which when solved gave our final unknown matrix:

\begin{equation} \label{eq24}
\mathbb{T}_X=\left(
\begin{array}{ccc}
    \frac{1}{a}(b\,\!\sin^{2}\phi+a\,\!cos^{2}\phi) &(a-b)\,\!\sin\,\!\phi\,\!\cos\,\!\phi & 0 \\
     (a-b)\,\!\sin\,\!\phi\,\!\cos\,\!\phi& \frac{1}{b}(a\,\!\sin^{2}\phi+b\,\!cos^{2}\phi) & 0 \\
     0 & 0 & f \\
  \end{array}
\right).
\end{equation}

Hence, we conclude that in the presence of strain in an arbitrary direction $\phi$ with respect to the crystallographic $x$-axis, the Raman tensor for the $A_1'$ mode is of the following form:

\begin{equation} \label{eq25}
\overline{\overline{\alpha}}({A_1'})_{\phi}=
\mathbb{T}_X
\left(
\begin{array}{ccc}
    a & 0 & 0 \\
    0 & b  & 0 \\
    0 & 0 & c \\
  \end{array}
\right).
\end{equation}

Therefore, the data for the $A_1'$ mode uncovers a way to determine the direction of strain by the angular dependence of the intensity of the $A'_1$ mode even in the absence of a knowledge of the crystallographic orientation. Once the strain direction is known using the information from the $A'_1$ mode, then it is possible to determine the crystallographic orientation using the information from the $E_\parallel'$ and $E_\perp'$ modes. Remarkably, all of this can be achieved by polarized Raman spectroscopy alone.

\section{Conclusions}

Using first-principles calculations, we have shown that a uniaxial in-plane strain breaks the symmetry of the doubly degenerate $E'$ mode in monolayer MoS$_2$, which splits into two orthogonal branches. In one mode, atoms vibrate parallel to the strain direction, denoted by $E_\parallel'$. In the other mode, atoms vibrate perpendicular to the strain direction, denoted by $E_\perp'$. Frequencies of these two modes exhibit different dependence on the applied strain: while the frequency of the $E_\perp'$ mode is almost unchanged, the $E_\parallel'$ mode blue-shifts (red-shifts) for a compressive (tensile) strain.

Due to the strain-induced anisotropy in MoS$_2$ lattice, the polarized Raman spectra of these two modes exhibit distinct angular dependence, as both discussed by numerical calculations and analytical derivations. For all three typical laser polarizations, we have presented a detailed analysis of the Raman intensity as a function of polarization angle. Combining the frequency shift and the angular dependence of the polarized Raman spectra of these modes, not only the magnitude of the strain can be probed, but also the strain direction could be determined. Therefore, the polarized Raman spectroscopy offers an efficient non-destructive way to probe the uniaxial strains in monolayer MoS$_2$. The polarized Raman spectra can be carried out in the conventional Raman measurement. Therefore, it is relatively straightforward without any additional complexity. In this context, the $E'$ mode is a fingerprint for the uniaxial strain in monolayer MoS$_2$. However, it requires a knowledge of the crystallographic orientation to determine the direction of an arbitrary strain.

We show that the $A_1'$ mode can provide an additional piece of information that helps provide a complete solution. The angular dependence of the intensity for the $A_1'$ mode is straightforward to further illustrate the strain direction, particularly with the parallel polarization setup. This information can be utilized to determine the direction of strain even in the absence of a knowledge of the crystallographic orientation. Once the strain direction is known, then the crystallographic orientation can also be determined using the information from the $E'$ mode. Remarkably, all of this can be achieved by polarized Raman spectroscopy. We also show that the strain results in a change in the ratio of the intensities of the $A_1'$ mode to the $E_\parallel'$ or $E_\perp'$ modes. These results provide an additional way to gauge the magnitude of the strain applied in monolayer MoS$_2$.

%
%

The conclusions of this work may be directly generalized to other semiconducting transition metal dichalcogenides, such as WSe$_2$, WS$_2$, and MoSe$_2$. The proposed method may also be applicable to other 2D materials, and therefore is general.

We thank Prof. Vincent Meunier for fruitful discussions. D. D. is supported by FCSM Fisher General Endowment and FCSM Undergraduate Research Committee at the Towson University. J.A.Y. acknowledges the Faculty Development and Research Committee grant (OSPR No. 140269) and the FCSM Fisher General Endowment at the Towson University. This work used the computing resources of Carver at NERSC.

\end{document}